\documentclass[11pt]{article}

\usepackage{amssymb,amsmath}
\usepackage[dvips]{graphicx}
\usepackage{verbatim}

\DeclareMathOperator{\poly}{poly}
\DeclareMathOperator{\Tr}{tr}
\DeclareMathOperator{\supp}{supp}
\DeclareMathOperator{\sign}{sign}
\DeclareMathOperator{\vol}{vol}

\newtheorem{thm}{Theorem}

\newtheorem{lem}[thm]{Lemma}
\newcommand{\vskipline}{\vskip 11pt}

\newcommand{\set}[1]{\lbrace #1 \rbrace}

\newcommand{\intersect}{\cap}

\newcommand{\ket}[1]{| #1 \rangle}

\newcommand{\norm}[1]{\lVert #1 \rVert}
\newcommand{\tensor}{\otimes}
\newcommand{\Tensor}{\bigotimes}

\newcommand{\RR}{\mathbb{R}}
\newcommand{\CC}{\mathbb{C}}
\newcommand{\PP}{\mathcal{P}}
\newcommand{\SSS}{\mathcal{S}}

\begin{document}

\title{Consistency of Local Density Matrices is QMA-complete}

\author{Yi-Kai Liu\\
Computer Science and Engineering\\
University of California, San Diego\\
\texttt{y9liu@cs.ucsd.edu}}

%

\date{June 16, 2006}

\maketitle

\begin{abstract}
Suppose we have an $n$-qubit system, and we are given a collection of local density matrices $\rho_1,\ldots,\rho_m$, where each $\rho_i$ describes a subset $C_i$ of the qubits.  We say that the $\rho_i$ are ``consistent'' if there exists some global state $\sigma$ (on all $n$ qubits) that matches each of the $\rho_i$ on the subsets $C_i$.  This generalizes the classical notion of the consistency of marginal probability distributions.  

We show that deciding the consistency of local density matrices is QMA-complete (where QMA is the quantum analogue of NP).  This gives an interesting example of a hard problem in QMA.  Our proof is somewhat unusual:  we give a Turing reduction from Local Hamiltonian, using a convex optimization algorithm by Bertsimas and Vempala, which is based on random sampling.  Unlike in the classical case, simple mapping reductions do not seem to work here.  
\end{abstract}


\section{Introduction}

Quantum mechanical systems exhibit many unusual phenomena, such as coherent superpositions and nonlocal entanglement.  It is interesting to compare this with the behavior of classical probabilistic systems.  In a classical system, such as a Markov chain or a graphical model, one may have correlations or dependencies among different parts of the system; in particular, local properties can affect the joint probability distribution of the entire system.  Many quantum systems have a similar flavor, though their behavior is more complicated.  In this paper, we investigate one problem of this kind, and its relationship to the complexity class QMA.  

First, consider a classical problem.  Suppose we have random variables $X_1,\ldots,X_n$, with some unknown joint distribution $D$, and we are given marginal distributions $D_1,\ldots,D_m$, where each $D_i$ describes a subset $C_i$ of the variables.  (We assume that the random variables $X_j$ take on values in some fixed finite set, and the subsets $C_i$ have size at most some constant $k$.)  Does there exist a joint distribution $D$ that matches each of the marginals $D_i$ on the subsets $C_i$?  If so, we say that the marginals $D_i$ are ``consistent.''  

Deciding the consistency of marginal distributions is NP-hard, by a straightforward reduction from 3-coloring.  (We are given a graph $G = (V,E)$.  For each vertex $v \in V$, construct a random variable $X_v$ which takes on values in $\set{r,g,b}$.  For each edge $(u,v) \in E$, specify that the marginal distribution of $X_u$ and $X_v$ must be uniform over the set $\set{r,g,b}^2 \setminus \set{rr,gg,bb}$.  These marginals are consistent iff $G$ is 3-colorable.)  

Now consider the generalization of this problem to quantum states.  (This problem was first suggested to me by Dorit Aharonov, in connection with the class QCMA \cite{A}.)  Suppose we have an $n$-qubit system, and we are given local density matrices $\rho_1,\ldots,\rho_m$, where each $\rho_i$ describes a subset $C_i$ of the qubits.  Does there exist a global state $\sigma$ on all $n$ qubits that matches each of the local states $\rho_i$ on the subsets $C_i$?  If so, we say that the local states $\rho_i$ are ``consistent.''  

We will show that this problem is QMA-complete, where QMA is the quantum analogue of NP.  QMA is the class of languages that have poly-time quantum verifiers, where the witness is allowed to be a quantum state.  QMA arises naturally in the study of quantum computation, and it also has a complete problem, Local Hamiltonian, which is a generalization of $k$-SAT \cite{KSV,AN}.  

Our result is interesting, because we only know of a few QMA-complete problems, and most of them look like universal models of quantum computation.  For instance, the fact that Local Hamiltonian is QMA-complete \cite{KSV,AN,KR,KKR,OT} is closely related to the fact that adiabatic quantum computation is equivalent to the standard quantum circuit model \cite{ADKLLR}.  Other QMA-complete problems such as Identity Check involve properties of quantum circuits \cite{JWB}.  The Consistency problem, however, does not seem to embody any particular model of quantum computation; this will become clearer when we present our reduction from Local Hamiltonian.  

Why are there so few QMA-complete problems, when there is such an astonishing variety of NP-complete problems?  The reason seems to be that the techniques used to show NP-hardness, such as mapping reductions using combinatorial gadgets, break down when we apply them to a ``quantum'' problem like Local Hamiltonian.  For instance, to reduce Local Hamiltonian to the Consistency problem, we would try to use local density matrices to ``simulate'' local Hamiltonians.  But we run into problems due to the presence of non-commuting matrices.  (In cases where quantum gadgets do work, such as \cite{KKR,OT}, they are much more subtle than classical gadgets.)  

Instead, our proof that the consistency problem is QMA-hard uses a randomized Turing reduction from Local Hamiltonian.  The basic idea is that Local Hamiltonian can be expressed as a convex program in polynomially many variables, which can be solved using convex optimization algorithms, given an oracle for the Consistency problem.  In particular, we use a class of convex optimization algorithms \cite{BV,KV,Vsurvey} which are based on random walks, and only require a membership oracle, rather than a separation oracle.  We also use a nifty representation of local density matrices in terms of the expectation values of Pauli matrices.  

Note that the Consistency problem has a rather different structure from Local Hamiltonian.  For instance, a local density matrix contains complete information about the local state of the system, whereas in many cases a local Hamiltonian only constrains the local state of the system to lie within a certain subspace.  

Finally, we remark that our reduction from Local Hamiltonian to Consistency preserves the ``neighborhood structure'' of the problem, in that the local density matrices act on the same subsets of qubits as the local Hamiltonians.  So, using the QMA-hardness results for 2-Local Hamiltonian \cite{KKR} and Local Hamiltonian on a 2-D square lattice \cite{OT}, we can immediately get QMA-hardness results for the corresponding special versions of the Consistency problem.  

We also mention some related work.  In \cite{BravyiVyalyi}, one considers the Common Eigenspace Problem, verifying the consistency of a set of eigenvalue equations $H_i \ket{\psi} = \lambda_i \ket{\psi}$, where the operators $H_i$ commute.  We do something similar, translating each local density matrix into constraints on the expectation values of Pauli matrices, though in our case the Pauli matrices do not commute.  Also, in \cite{Bravyi}, one considers a quantum analogue of 2-SAT, where we seek a state $\ket{\psi}$ whose local density matrices have support on prescribed subspaces.  However, this problem is more closely related to Local Hamiltonian than to Consistency, since the constraints can be written in the form $\Pi_i \ket{\psi} = 0$ where the $\Pi_i$ are local projectors.


\section{Preliminaries}

\subsection{Density Matrices}

A quantum state of an $n$-qubit system is represented by a density matrix, which is a $2^n \times 2^n$ positive semidefinite matrix with trace 1.  A classical joint probability distribution on $n$ bits is a special case, where the density matrix is diagonal, and the diagonal entries are the probabilities of the $2^n$ possible outcomes.  A subset of qubits $C$ is described by a reduced density matrix, which is obtained by taking the partial trace over the qubits not in $C$.  This is analogous to a marginal distribution, which is obtained by summing over some of the variables.  

We measure the difference between two quantum states using the $L_1$ matrix norm, $\norm{\rho-\sigma}_1 = \Tr|\rho-\sigma|$.  Note that this is also called the trace or statistical distance (when normalized by a factor of 1/2).  


Let $X$, $Y$ and $Z$ denote the Pauli matrices for a single qubit, and define $\PP = \set{I,X,Y,Z}$.  We can construct $n$-qubit Pauli matrices by taking tensor products $P = P_1 \tensor \cdots \tensor P_n \in \PP^{\tensor n}$.  Any $2^n$-dimensional Hermitian matrix can be written as a real linear combination of $n$-qubit Pauli matrices.  Furthermore, the $n$-qubit Pauli matrices are orthogonal with respect to the Hilbert-Schmidt inner product:  $\Tr(P^\dagger Q) = 2^n$ if $P = Q$, and 0 otherwise.  So, if $\sigma$ is an $n$-qubit state, we can write it in the form 
\[
\sigma = \frac{1}{2^n} \sum_{P \in \PP^{\tensor n}} \alpha_P P, 
\]
where the coefficients are uniquely determined by $\alpha_P = \Tr(P\sigma)$; note that these are the expectation values of the Pauli matrices $P$.  This application of the Pauli matrices is closely related to quantum state tomography.


\subsection{QMA and the Local Hamiltonian Problem}

The class QMA, or ``Quantum Merlin-Arthur,'' is defined as follows \cite{KSV,AN}:  a language $L$ is in QMA if there exists a poly-time quantum verifier $V$ and a polynomial $p$ such that 
\begin{itemize}
\item If $x \in L$, then there exists a quantum state $\rho$ on $p(|x|)$ qubits such that $V(x,\rho)$ accepts with probability $\geq 2/3$.  
\item If $x \notin L$, then for all quantum states $\rho$ on $p(|x|)$ qubits, $V(x,\rho)$ accepts with probability $\leq 1/3$.  
\end{itemize}
(Here, $|x|$ denotes the length of the string $x$.)  This is similar to the definition of NP, except that the witness is allowed to be a quantum state, and the verifier is a quantum circuit with bounded error probability.  

The Local Hamiltonian problem is defined as follows:  
\begin{quote}
Consider a system of $n$ qubits.  We are given a Hamiltonian $H = H_1+\cdots+H_m$, where each $H_i$ acts on a subset of qubits $C_i \subseteq \set{1,\ldots,n}$.  The $H_i$ are Hermitian matrices, with eigenvalues in some fixed interval (for instance $[0,1]$), and each matrix entry is specified with $\poly(n)$ bits of precision.  Also, $m \leq \poly(n)$, and each subset $C_i$ has size $|C_i| \leq k$, for some constant $k$.  

In addition, we are given two real numbers $a$ and $b$ (specified with $\poly(n)$ bits of precision) such that $b-a \geq 1/\poly(n)$.  

The problem is to distinguish between the following two cases:  
\begin{itemize}
\item If $H$ has an eigenvalue that is $\leq a$, output ``YES.''  
\item If all the eigenvalues of $H$ are $\geq b$, output ``NO.''  
\end{itemize}
\end{quote}
Kitaev showed that Local Hamiltonian is in QMA, and the case of $k=5$ is QMA-hard \cite{KSV,AN}.  With greater effort, one can show that Local Hamiltonian with $k=2$ is also QMA-hard \cite{KR,KKR}.


\subsection{Convex Programming}

Consider the following version of convex programming:  
\begin{verse}
Let $K \subseteq \RR^n$ be a convex set, specified by a membership oracle $O_K$.\\
Let $f:\: K \rightarrow \RR$ be a convex function, which is efficiently computable.\\
Find some $x \in K$ that minimizes $f(x)$.
\end{verse}
Note that the membership oracle $O_K$ is not as powerful as a separation oracle.  We would like to solve this problem with precision $\varepsilon$; that is, we want to find some $x$ that lies within distance $\varepsilon$ of an optimal solution $x^*$.  

We can solve this problem in time $\poly(n,\log(1/\varepsilon))$, using an algorithm by Bertsimas and Vempala which is based on random sampling \cite{BV,Vsurvey}.  Actually, for our purposes we only need to solve the special case where $f$ is a linear function; for this case, we can use a slightly faster simulated annealing algorithm \cite{KV}, or an algorithm based on the shallow-cut Ellipsoid method \cite{GLS}.  But for simplicity we will stick with the Bertsimas and Vempala algorithm.  

\begin{thm}\label{thm-BV}
(Bertsimas and Vempala) Consider the convex program described above.  Suppose $K$ is contained in a ball of radius $R$ centered at the origin.  Also, suppose we are given a point $y$, such that the ball of radius $r$ around $y$ is contained in $K$.  Then this problem can be solved in time $\poly(n,L)$, where $L = \log(R/r)$.  
\end{thm}


\section{Consistency of Local Density Matrices}

We define the Consistency problem as follows \cite{A}:  
\begin{quote}
Consider a system of $n$ qubits.  We are given a collection of local density matrices $\rho_1,\ldots,\rho_m$, where each $\rho_i$ acts on a subset of qubits $C_i \subseteq \set{1,\ldots,n}$.  Each matrix entry is specified with $\poly(n)$ bits of precision.  Also, $m \leq \poly(n)$, and each subset $C_i$ has size $|C_i| \leq k$, for some constant $k$.  

In addition, we are given a real number $\beta$ (specified with $\poly(n)$ bits of precision) such that $\beta \geq 1/\poly(n)$.  

The problem is to distinguish between the following two cases:  
\begin{itemize}
\item There exists an $n$-qubit state $\sigma$ such that, for all $i$, $\norm{\Tr_{\set{1,\ldots,n}-C_i}(\sigma) - \rho_i}_1 = 0$.  In this case, output ``YES.''  
\item For all $n$-qubit states $\sigma$, there exists some $i$ such that $\norm{\Tr_{\set{1,\ldots,n}-C_i}(\sigma) - \rho_i}_1 \geq \beta$.  In this case, output ``NO.''  
\end{itemize}
\end{quote}

\begin{thm}\label{thm-QMA}
Consistency is in QMA.  
\end{thm}

\noindent
Proof:  The basic idea is as follows.  Given a witness state $\sigma$, the verifier will pick a subset $C_i$, and perform measurements to compare $\sigma$ (on the subset $C_i$) to $\rho_i$.  There is a complication, however, because the verifier requires many independent copies of the witness $\sigma$, and the prover might try to cheat using entanglement among the different copies.  One can deal with this problem using a Markov argument.  

A concise way to present this result is to use QMA+, an alternate characterization of QMA \cite{AR}.  We construct a QMA+ ``super-verifier'' for Consistency:  
\begin{quote}
Choose $i \in \set{1,\ldots,m}$ at random.  Choose a Pauli matrix $Q \in \PP^{\tensor |C_i|}$ (acting on the subset $C_i$), at random.  

Output a quantum circuit $M$ which does the following:  measure the observable $Q \tensor I$ on the state $\sigma$; if the measurement outcome is $+1$, accept; if it is $-1$, reject.  [So the circuit accepts with probability $\tfrac{1}{2} + \tfrac{1}{2} \Tr((Q \tensor I) \sigma)$.]  

Set the target probability to be $r = \tfrac{1}{2} + \tfrac{1}{2} \Tr(Q \rho_i)$, and set the error threshold to be $s = \tfrac{1}{2} (\beta/4^{|C_i|})$.  [Note that $s \geq \tfrac{1}{2} (\beta/4^k) = 1/\poly(n)$.]  
\end{quote}
To see why this works, write the difference between $\sigma$ and $\rho_i$ on subset $C_i$ as follows:  
\[
\Tr_{\set{1,\ldots,n}-C_i}(\sigma) - \rho_i
 = \frac{1}{2^{|C_i|}} \sum_{Q \in \PP^{\tensor |C_i|}} \gamma_Q Q, 
\]
where $\gamma_Q = \Tr((Q \tensor I) \sigma) - \Tr(Q \rho_i)$.  On a ``yes'' instance, we always have $\gamma_Q = 0$, so the circuit $M$ accepts with probability $r$.  On a ``no'' instance, there must be some choice of $C_i$ and $Q$ such that $|\gamma_Q| \geq \beta/4^{|C_i|}$; to see this, use the triangle inequality to write:  
\[
\beta \leq \norm{\Tr_{\set{1,\ldots,n}-C_i}(\sigma) - \rho_i}_1
 \leq \sum_{Q \in \PP^{\tensor |C_i|}} |\gamma_Q|.  
\]
In this case, the circuit $M$ accepts with probability that differs from $r$ by at least $s$.  Moreover, the probability that the super-verifier will pick this $C_i$ and $Q$ is at least $1/4^km = 1/\poly(n)$.  $\square$


\section{Consistency is QMA-hard}

\begin{thm}\label{thm-QMA-hard}
Consistency is QMA-hard, via a poly-time randomized Turing reduction from Local Hamiltonian.  Furthermore, the reduction uses the same value of $k$ for both problems, so we get that Consistency with $k=2$ is QMA-hard.  
\end{thm}
We begin by discussing the basic idea of the proof, and the complications that arise.  We then describe the actual reduction from Local Hamiltonian to Consistency, and finally we deal with some issues of numerical precision.


\subsection{The Basic Idea}

Say we are given a local Hamiltonian $H = H_1+\cdots+H_m$, where $H_i$ acts on the subset $C_i$.  Consider the following convex program:  
\begin{verse}
Let $\rho$ be any $2^n \times 2^n$ complex matrix.\\
Find some $\rho$ that minimizes $\Tr(H\rho)$,\\
such that $\rho \succeq 0$ and $\Tr(\rho) = 1$.
\end{verse}
It is easy to see that $H$ has an eigenvalue $\leq \gamma$ if and only if the convex program achieves $\Tr(H\rho) \leq \gamma$ for some $\rho$.  (Note that, although the convex program allows mixed states $\rho$, the optimal $\rho$ can always be chosen to be a pure state.)  Unfortunately, this convex program has $4^n$ variables, which makes it unwieldy.  

We now construct another convex program, which is equivalent to the previous one, but has only a polynomial number of variables:  
\begin{verse}
Let $\rho_1,\ldots,\rho_m$ be complex matrices, where $\rho_i$ has size $2^{|C_i|} \times 2^{|C_i|}$.\\
(We interpret each $\rho_i$ as the reduced density matrix for the subset $C_i$.)\\
Find some $\rho_1,\ldots,\rho_m$ that minimize $\Tr(H_1\rho_1) + \cdots + \Tr(H_m\rho_m)$,\\
such that each $\rho_i$ satisfies $\rho_i \succeq 0$ and $\Tr(\rho_i) = 1$,\\
and $\rho_1,\ldots,\rho_m$ are consistent.
\end{verse}
Note that the set of feasible solutions is indeed convex:  if $\rho_1,\ldots,\rho_m$ are consistent, and $\rho'_1,\ldots,\rho'_m$ are consistent, then any convex combination $\rho''_i = q\rho_i+(1-q)\rho'_i$ ($i=1,\ldots,m$) is also consistent.  

Observe that the optimal value of this convex program is equal to the optimal value of the previous convex program; this is because, if $\rho_1,\ldots,\rho_m$ are consistent with some $n$-qubit state $\sigma$, then $\Tr(H\sigma) = \Tr(H_1\rho_1) + \cdots + \Tr(H_m\rho_m)$.  Also, note that the number of variables in this convex program is $\sum_{i=1}^m 4^{|C_i|} \leq 4^k m \leq \poly(n)$.  

This convex program has a ``consistency'' constraint, which we do not know how to evaluate.  But if we have an oracle for the Consistency problem, then we can solve this convex program, using the algorithm of Bertsimas and Vempala.  To make this work, we will have to find a suitable representation for the set of feasible solutions, 
\[
K = \set{(\rho_1,\ldots,\rho_m) \text{ which are consistent}}.  
\]
Also, we will have to address some questions about the accuracy of the Consistency oracle, i.e., how well does it approximate $K$, and how does this affect the Bertsimas-Vempala algorithm.  

We could represent each element $(\rho_1,\ldots,\rho_m) \in K$ by writing down the matrix entries for the $\rho_i$; then we could view $K$ as a subset of $\CC^d$, where $d = \sum_{i=1}^m 4^{|C_i|}$.  But this straightforward approach runs into some trouble.  Observe that the matrix entries must satisfy some algebraic constraints:  each $\rho_i$ must be Hermitian, $(\rho_i)^\dagger = \rho_i$; and $\rho_i$ and $\rho_j$ must agree on their intersection $C_i \intersect C_j$, that is, $\Tr_{C_i - (C_i \intersect C_j)}(\rho_i) = \Tr_{C_j - (C_i \intersect C_j)}(\rho_j)$.  Because of these constraints, the set $K$ actually lies in a lower-dimensional subspace of $\CC^d$.  We would need to characterize this subspace, before we can apply the Bertsimas-Vempala algorithm.  We can avoid this problem by switching to a different representation for the set $K$.


\subsection{The Actual Reduction}

We will represent each element of $K$ using the expectation values of the ``local'' Pauli matrices on the subsets $C_1,\ldots,C_m$.  These local Pauli matrices form a basis for the space of all local Hamiltonians (acting on the subsets $C_i$).  For an $n$-qubit state $\sigma$, knowing the expectation values of these Pauli matrices is equivalent to knowing the projection of $\sigma$ onto this subspace; and this is equivalent to knowing the local density matrices of $\sigma$.  

First, some notation.  Let $P$ be an $n$-qubit Pauli matrix, and define the ``support'' of $P$ be the set of qubits on which $P$ acts nontrivially; that is, $\supp(P) = \set{i \:|\: P_i \neq I}$.  Also, for any subset of qubits $C$, define the ``restriction'' of $P$ to $C$, $P|C = \Tensor_{i \in C} P_i$.  

Define $\SSS$ to be the set of ``local'' Pauli matrices:  
\[
\SSS = \set{P \in \PP^{\tensor n} \:|\: \supp(P) \subseteq C_i \text{ for some } i} 
- \set{I}, 
\]
where we excluded the identity matrix $I$ because its expectation value is always 1.  Also let $d = |\SSS|$, and note that $d \leq 4^k m - 1 \leq \poly(n)$.  

For each $P \in \SSS$, let $\alpha_P$ be the corresponding expectation value; and let $(\alpha_P)_{P \in \SSS}$ denote the collection of these $\alpha_P$.  Also, let $\alpha_I = 1$.  We define the set $K'$ to be 
\[
K' = \set{(\alpha_P)_{P \in \SSS} \text{ which are consistent}}, 
\]
where we say the $\alpha_P$ are ``consistent'' if there exists an $n$-qubit state $\sigma$ such that for all $P \in \SSS$, $\alpha_P = \Tr(P\sigma)$.  Note that $K'$ is a subset of $\RR^d$.  Also, clearly $K'$ is convex.  

\begin{lem}\label{lemma-K-represent}
There is a linear bijection between $K$ and $K'$.  
\end{lem}

\noindent
Proof:  Given some $(\rho_1,\ldots,\rho_m) \in K$, we can construct $(\alpha_P)_{P \in \SSS} \in K'$ as follows:  
\begin{quote}
For each $P \in \SSS$:  We know that $\supp(P) \subseteq C_i$ for some $i$.  So we can write $P$ in the form $P = (P|C_i) \tensor I$.  Then we set $\alpha_P = \Tr((P|C_i) \rho_i)$.  
\end{quote}
If the $\rho_i$ are consistent with some $n$-qubit state $\sigma$, then the $\alpha_P$ are also consistent with $\sigma$; to see this, write $\alpha_P = \Tr((P|C_i) \rho_i) = \Tr(P\sigma)$.  (Note that in the case where $\supp(P) \subseteq C_i \intersect C_j$, it makes no difference whether we pick $i$ or $j$ in the above procedure, because $\rho_i$ and $\rho_j$ yield the same reduced density matrix on $C_i \intersect C_j$.)  

Going in the opposite direction, given some $(\alpha_P)_{P \in \SSS} \in K'$, we can construct $(\rho_1,\ldots,\rho_m) \in K$ as follows:  
\begin{quote}
For $i = 1,\ldots,m$:  We construct $\rho_i$ by using the $\alpha_P$ for all $P$ with $\supp(P) \subseteq C_i$.  Note that we can write $P$ in the form $P = (P|C_i) \tensor I$.  We set 
\[
\rho_i = \frac{1}{2^{|C_i|}} \sum_{P \::\: \supp(P) \subseteq C_i} \alpha_P (P|C_i).  
\]
\end{quote}
If the $\alpha_P$ are consistent with some $n$-qubit state $\sigma$, then the $\rho_i$ are also consistent with $\sigma$; to see this, write $\sigma$ in terms of the $\alpha_P$, where we now include the expectation values $\alpha_P = \Tr(P\sigma)$ for all $P \in \PP^{\tensor n}$, 
\[
\sigma = \frac{1}{2^n} \sum_{P \in \PP^{\tensor n}} \alpha_P P; 
\]
note that when we trace out the qubits not in $C_i$, we get that $\Tr_{\set{1,\ldots,n}-C_i}(P)$ equals $2^{n-|C_i|} (P|C_i)$ if $\supp(P) \subseteq C_i$, and 0 otherwise; thus we have 
\[
\Tr_{\set{1,\ldots,n}-C_i}(\sigma)
 = \frac{1}{2^{|C_i|}} \sum_{P \::\: \supp(P) \subseteq C_i} \alpha_P (P|C_i)
 = \rho_i.  
\]

Finally, observe that these maps (between $K$ and $K'$) are linear, and they are inverses of each other.  $\square$

\vskipline

So we can restate our convex program, using the set $K'$ instead:  
\begin{verse}
Let $\alpha_P$ (for $P \in \SSS$) be real numbers.\\
Find some $\alpha_P$ that minimize 
\[
\sum_{i=1}^m \frac{1}{2^{|C_i|}} 
\sum_{P \::\: \supp(P) \subseteq C_i} \alpha_P \Tr(H_i (P|C_i)), 
\]\\
such that $(\alpha_P)_{P \in \SSS} \in K'$ (i.e., the $\alpha_P$ are consistent).
\end{verse}

\begin{lem}\label{lemma-convex-prog}
The optimal value of this convex program is equal to the smallest eigenvalue of the local Hamiltonian $H = H_1+\cdots+H_m$.  
\end{lem}

\noindent
Proof:  This follows from the remarks in the previous section, and Lemma \ref{lemma-K-represent}.  $\square$

\vskipline

Next, we prove some bounds on the geometry of the set $K' \subseteq \RR^d$.  

\begin{lem}\label{lemma-R}
$K'$ is contained in a ball of radius $R = \sqrt{d}$ centered at the origin.  
\end{lem}

\noindent
Proof:  Suppose $(\alpha_P)_{P \in \SSS} \in K'$, and say it is consistent with some state $\sigma$.  Since $\alpha_P = \Tr(P\sigma)$, it follows that $-1 \leq \alpha_P \leq 1$, which implies the result.  $\square$

\vskipline

\begin{lem}\label{lemma-r}
The ball of radius $r = 1/\sqrt{d}$ around the origin is contained in $K'$.  
\end{lem}

\noindent
Proof:  Let $(\alpha_P)_{P \in \SSS}$ be any vector in $\RR^d$ of length at most $1/\sqrt{d}$.  By the Cauchy-Schwartz inequality, $\sum_{P \in \SSS} |\alpha_P| \leq 1$; let $p = \sum_{P \in \SSS} |\alpha_P|$.  Now define $\sigma = (1/2^n) (I + \sum_{P \in \SSS} \alpha_P P)$.  This is a legal density matrix, because it can be written as 
\[
\begin{split}
\sigma
 &= \frac{1}{2^n} \Bigl( (1-p)I + \sum_{P \in \SSS} (|\alpha_P|I + \alpha_P P) \Bigr)\\
 &= (1-p) \frac{I}{2^n} + \sum_{P \in \SSS} |\alpha_P| \frac{I+\sign(\alpha_P)P}{2^n}, 
\end{split}
\]
which is (with probability $1-p$) the fully mixed state, and (with probability $|\alpha_P|$, for $P \in \SSS$) the mixture of all eigenstates of $P$ with eigenvalue $\sign(\alpha_P)$.  Furthermore, the $\alpha_P$ are consistent with $\sigma$; thus we conclude that $(\alpha_P)_{P \in \SSS} \in K'$.  $\square$


\subsection{Numerical Precision}

First, we will show that the Consistency oracle gives a good approximation to the set $K'$.  We start by defining a new problem, Consistency$'$, using the expectation values $\alpha_P$ of the local Pauli matrices $P \in \SSS$ (similar to the definition of $K'$):  
\begin{quote}
As in the original Consistency problem, we have an $n$-qubit system, and subsets $C_1,\ldots,C_m$, with $|C_i| \leq k$.  But instead of the local density matrices $\rho_1,\ldots,\rho_m$, we are given real numbers $\alpha_P$ for all $P \in \SSS$.  Each $\alpha_P$ is specified with $\poly(n)$ bits of precision.  

In addition, we are given a real number $\beta'$ (specified with $\poly(n)$ bits of precision) such that $\beta' \geq 1/\poly(n)$.  

The problem is to distinguish between the following two cases:  
\begin{itemize}
\item There exists an $n$-qubit state $\sigma$ such that, for all $P \in \SSS$, $\Tr(P\sigma) = \alpha_P$.  In this case, output ``YES.''  
\item For all $n$-qubit states $\sigma$, $\bigl( \sum_{P \in \SSS} (\Tr(P\sigma)-\alpha_P)^2 \bigr)^{1/2} \geq \beta'$.  In this case, output ``NO.''  
\end{itemize}
\end{quote}

\begin{lem}\label{lemma-consistency-prime}
There is a poly-time mapping reduction from Consistency$'$ to Consistency.  
\end{lem}

\noindent
Proof:  The reduction is as follows:  Use the $\alpha_P$ to construct $\rho_1,\ldots,\rho_m$ as described in Lemma \ref{lemma-K-represent}.  Set $\beta = \beta'/\sqrt{d}$, where $d = |\SSS|$.  

Clearly, a ``YES'' instance of Consistency$'$ maps to a ``YES'' instance of Consistency.  Now suppose we have a ``NO'' instance of Consistency$'$.  Then for all $\sigma$ there is some $P \in \SSS$ such that $|\Tr(P\sigma)-\alpha_P| \geq \beta'/\sqrt{d}$.  We know that $\supp(P) \subseteq C_i$ for some $i$, so we can write $P = \tilde{P} \tensor I$ where $\tilde{P}$ acts on the subset $C_i$.  Also, let $\tilde{\sigma} = \Tr_{\set{1,\ldots,n}-C_i}(\sigma)$.  Then we have 
\[
\bigl| \Tr(\tilde{P} \tilde{\sigma})
 - \Tr(\tilde{P} \rho_i) \bigr| \geq \beta'/\sqrt{d}.  
\]

We will use $\tilde{P}$ to construct a POVM measurement that distinguishes between $\tilde{\sigma}$ and $\rho_i$.  Since the eigenvalues of $\tilde{P}$ are all $\pm 1$, we can write $\tilde{P} = \Pi_1-\Pi_2$, where $\Pi_1$ and $\Pi_2$ are projectors on orthogonal subspaces, and $\Pi_1+\Pi_2 = I$.  Thus we can use $\set{\Pi_1,\Pi_2}$ as a POVM.  For the state $\tilde{\sigma}$, let $s_j$ be the probability of measuring $j$ (for $j = 1,2$); and for the state $\rho_i$, let $r_j$ be the probability of measuring $j$ (for $j = 1,2$).  

Then we have 
\[
\bigl| \Tr(\tilde{P} \tilde{\sigma})
 - \Tr(\tilde{P} \rho_i) \bigr|
 = |(s_1-s_2) - (r_1-r_2)|
 = 2|s_1-r_1|.  
\]
Observe that the $\ell_1$ distance between $s$ and $r$ is $\norm{s-r}_1 = |s_1-r_1| + |s_2-r_2| = 2|s_1-r_1|$.  Also, $\norm{s-r}_1$ is a lower bound for the $L_1$ (matrix) distance between $\tilde{\sigma}$ and $\rho_i$.  So we have 
\[
\norm{\tilde{\sigma} - \rho_i}_1
 \geq \norm{s-r}_1 \geq \beta'/\sqrt{d} = \beta.  
\]
Thus we have a ``NO'' instance of Consistency.  $\square$

\vskipline

Next, we will show that the Bertsimas-Vempala algorithm succeeds in solving our convex program, even when the oracle for the set $K'$ is slightly inaccurate.  (The shallow-cut ellipsoid method would also work, see \cite{GLS} for details.)  We will make some general remarks about the algorithm, and then show that it works for our specific problem.  

For our purposes, we only need to solve a simpler problem, deciding the feasibility of a convex program:  
\begin{verse}
As before, let $K$ be a convex set, and let $f$ be a convex function.\\
Given some $t \in \RR$, does there exist a point $x \in K$ such that $f(x) \leq t$?
\end{verse}

The Bertsimas-Vempala algorithm is built around a subroutine that solves the feasibility problem \cite{BV}.  The basic idea is as follows:  
\begin{verse}
Let $P$ be the set $K$.\\
Randomly sample some points from $P$, and compute an approximate centroid of $P$; call this point $z$.\\
If $f(z) \leq t$, stop and return true.\\
Compute $\nabla f(z)$, and use this to cut out a portion of the set $P$.\footnote{Specifically, we can deduce a hyperplane that separates $z$ from the set $\set{x \:|\: f(x) \leq t}$.  Then we take the intersection of $P$ with the half-space that does not contain $z$.}\\
Repeat the procedure starting from line 2.  If $P$ gets too small, stop and return false.
\end{verse}

The critical step is to sample random points from the set $P$.  (Note that $P$ is convex, and we have a membership oracle for $P$.)  One way is to do a random walk known as the ``ball walk'':  
\begin{quote}
Pick a point $y$ uniformly at random in the ball of radius $\delta$ centered at the current position $x$.  If $y \in P$, then move to $y$, otherwise stay at $x$.  Repeat.  
\end{quote}

The points where the membership oracle makes mistakes all lie close to the boundary of $P$; call this the ``boundary layer'' $P_b$.  Intuitively, if the boundary layer is thin, it should not have much effect on the random walk.  Using an argument by Lov\'asz and Simonovits \cite{LS93}, one can prove (omitting some details):  
\begin{lem}\label{lemma-LS}
For any polynomial $p$, there exists a polynomial $q$ such that, if we run the $n$-dimensional ball walk for at most $p(n)$ steps, and $\vol(P_b)/\vol(P) \leq 1/q(n)$, then with probability $2/3$ we will never enter the region $P_b$.  
\end{lem}
So, if we can show that the boundary layer is small compared to the total volume of $P$, then our algorithm will work fine.  (As long as the random walk does not enter the boundary layer, the algorithm will perform exactly as if it had access to a perfect membership oracle.)  

Finally, there may still be errors due to finite numerical precision---using $n$ bits of precision, we have errors of size $2^{-n}$.  This will not be a problem for us, since we only need accuracy of $1/\poly(n)$.  

We are now ready to prove that the Bertsimas-Vempala algorithm works for our specific problem:  

\vskipline

\noindent
Proof of Theorem \ref{thm-QMA-hard}:  Given an instance of Local Hamiltonian, use Lemma \ref{lemma-convex-prog} to express it as a convex program over the set $K' \subseteq \RR^d$, where $d \leq \poly(n)$.  Let $f$ denote the objective function, and set $t = (a+b)/2$.  By Lemma \ref{lemma-consistency-prime}, we can assume we have an oracle for Consistency$'$, which approximates the set $K'$ with error $\beta'$, for any $\beta' \geq 1/\poly(n)$.  Then use the Bertsimas-Vempala algorithm to solve the following problem:  does there exist a solution $\alpha \in K'$ such that $f(\alpha) \leq t$?  

Recall that the objective function 
\[
f(\alpha) = \sum_{i=1}^m \frac{1}{2^{|C_i|}} 
\sum_{P \::\: \supp(P) \subseteq C_i} \alpha_P \Tr(H_i P)
\]
is linear.  We claim that its derivatives in all directions are at most $4^k m \leq \poly(n)$.  To see this, note that there are at most $4^k m$ terms in the sum, and for each term, we have 
\[
|\Tr(H_i P)| \leq \Tr(|H_i P|)
 \leq \norm{H_i}_2 \norm{P}_2 \leq 2^{|C_i|}, 
\]
using the Cauchy-Schwartz inequality for the $L_2$ matrix norm \cite{Bhatia}, and the fact that the eigenvalues of $H_i$ lie in the interval $[0,1]$, while the eigenvalues of $P$ are $\pm 1$.  

Now suppose we have a ``YES'' instance of Local Hamiltonian.  Then there exists some $\alpha^* \in K'$ such that $f(\alpha^*) \leq a$.  We claim that the set $\set{\alpha \in K' \:|\: f(\alpha) \leq t}$ contains a ball of radius $\delta \geq 1/\poly(n)$.  To see this, let $\sigma^*$ be the density matrix which corresponds to $\alpha^*$.  Perturb $\sigma^*$ by mixing it with the state $I/2^n$, then add a small contribution of each of the Pauli matrices $P \in \SSS$.  This generates a ball contained in $K'$.  Moreover, this ball can have radius $\delta \geq 1/\poly(n)$ and still satisfy the condition $f(\alpha) \leq t$; this is because $f$ does not vary too quickly, and there is a gap between $a$ and $t$.  

So, in the Bertsimas-Vempala algorithm, the set $P$ always contains a ball of radius $\delta$.  Now set the error threshold for the membership oracle to be $\beta' \leq \delta/d^c$, for some constant $c$ (to be specified later).  We will show that the boundary layer $P_b$ is small compared to the total volume of $P$.  Define $P^+$ to be the set $P$ expanded by an amount $\beta'$, that is, $P^+ = P + \beta' B$, where $B$ is the unit ball.  We have that 
\[
P^+ \subseteq P + (\beta'/\delta) P = (1 + \beta'/\delta) P, 
\]
where the equality holds because $P$ is convex.  This implies that 
\[
\vol(P^+)
 \leq (1 + \beta'/\delta)^d \vol(P)
 \leq (1 + 2/d^{c-1}) \vol(P).  
\]
So we can conclude that $\vol(P_b) \leq \vol(P^+) - \vol(P) \leq (2/d^{c-1}) \vol(P)$.  

Finally, we choose the constant $c$.  We will show, later, that the Bertsimas-Vempala algorithm runs in time $\poly(n)$.  (This upper bound will not depend on the oracle error $\beta'$, hence it will not depend on $c$.)  We then choose $c$ sufficiently large so that Lemma \ref{lemma-LS} applies.  Therefore, the algorithm will work correctly in this case.  

Now suppose we have a ``NO'' instance of Local Hamiltonian.  Then for all $\alpha \in K'$, $f(\alpha) \geq b$.  In addition, there is some $\delta \geq 1/\poly(n)$ such that, for all $\alpha$ within distance $\delta$ of $K'$, $f(\alpha) > t$; this is because $f$ does not vary too quickly, and there is a gap between $b$ and $t$.  

Set the error threshold for the membership oracle to be $\beta' \leq \delta$.  Then the set $\set{\alpha \in K' \:|\: f(\alpha) \leq t}$ is empty, even when the membership oracle makes mistakes.  So the Bertsimas-Vempala algorithm will work correctly in this case.  

Finally, we claim that the Bertsimas-Vempala algorithm runs in time polynomial in $n$.  This follows from Theorem \ref{thm-BV} and Lemmas \ref{lemma-R} and \ref{lemma-r}; note that $L = \log(R/r) = \log(\poly(n)) = O(\log n)$.  $\square$


\section{Discussion}

Consistency of local density matrices is an interesting problem that gives some new insight into the class QMA.  The reduction from Local Hamiltonian is nontrivial, and in that sense, Consistency seems to be an easier problem to deal with.  One direction for future work is to try to find additional QMA-complete problems by giving reductions from Consistency (rather than from Local Hamiltonian).  

Another question is whether Consistency remains QMA-hard under mapping reductions.  We mention that we can build zero-knowledge proof systems for Consistency \cite{Liu}, using techniques developed by Watrous \cite{Wat}.  If we could show that Consistency is QMA-hard under mapping reductions, then we could get zero-knowledge proof systems for any language in QMA.  

\vskipline

\noindent
\textit{Acknowledgements:}  Thanks to Dorit Aharonov for suggesting this problem and pointing out an error in a previous version of the paper; thanks also to Russell Impagliazzo and the anonymous reviewers for their helpful comments.  Supported by an ARO/NSA Quantum Computing Graduate Research Fellowship.



\begin{thebibliography}{99}
\bibitem{A} D. Aharonov, private communication, 2004.  
\bibitem{KSV} A.Yu. Kitaev, A.H. Shen and M.N. Vyalyi, \textit{Classical and Quantum Computation}, AMS, 2002.  
\bibitem{AN} D. Aharonov and T. Naveh, ``Quantum NP - A Survey,'' Arxiv: quant-ph/0210077.  
\bibitem{KR} J. Kempe and O. Regev, ``3-Local Hamiltonian is QMA-complete,'' Quantum Info. and Comput., Vol.3(3), pp.258-264, 2003, Arxiv: quant-ph/0302079.  
\bibitem{KKR} J. Kempe, A. Kitaev and O. Regev, ``The Complexity of the Local Hamiltonian Problem,'' FSTTCS 2004, pp.372-383, Arxiv: quant-ph/0406180.  
\bibitem{OT} R. Oliveira and B.M. Terhal, ``The complexity of quantum spin systems on a two-dimensional square lattice,'' Arxiv: quant-ph/0504050.  
\bibitem{ADKLLR} D. Aharonov, W. van Dam, J. Kempe, Z. Landau, S. Lloyd and O. Regev, ``Adiabatic Quantum Computation is Equivalent to Standard Quantum Computation,'' FOCS 2004, pp.42-51, Arxiv: quant-ph/0405098.  
\bibitem{JWB} D. Janzing, P. Wocjan and T. Beth, ``Identity check is QMA-complete,'' Arxiv: quant-ph/0305050.  
\bibitem{BV} D. Bertsimas and S. Vempala, ``Solving Convex Programs by Random Walks,'' Journal of the ACM 51 (4) pp.540-556 (2004).  
\bibitem{KV} A. Kalai and S. Vempala, ``Convex Optimization by Simulated Annealing,'' preprint, 2004.  
\bibitem{Vsurvey} S. Vempala, ``Geometric Random Walks:  A Survey,'' MSRI volume on Combinatorial and Computational Geometry, 2005.  
\bibitem{GLS} M. Gr\"otschel, L. Lov\'asz and A. Schrijver, \textit{Geometric Algorithms and Combinatorial Optimization}, Springer, 1988.  
\bibitem{BravyiVyalyi} S. Bravyi and M. Vyalyi, ``Commutative version of the local Hamiltonian problem and common eigenspace problem,'' Quantum Info. and Comput., Vol.5, No.3 (2005), pp.187-215, Arxiv: quant-ph/0308021.  
\bibitem{Bravyi} S. Bravyi, ``Efficient algorithm for a quantum analogue of 2-SAT,'' Arxiv: quant-ph/0602108.  
\bibitem{AR} D. Aharonov and O. Regev, ``A Lattice Problem in Quantum NP,'' FOCS 2003, pp.210-219, Arxiv: quant-ph/0307220.  
\bibitem{LS93} L. Lov\'asz and M. Simonovits, ``Random Walks in a Convex Body and an Improved Volume Algorithm,'' Random Structures and Algorithms, Vol.4, No.4 (1993).  
\bibitem{Bhatia} R. Bhatia, \textit{Matrix Analysis}, Springer, 1997.  
\bibitem{Liu} Y.-K. Liu, in preparation.  
\bibitem{Wat} J. Watrous, ``Zero-knowledge against quantum attacks,'' Arxiv: quant-ph/0511020.  
\end{thebibliography}
\end{document}